\documentstyle[12pt,epsf]{article}

\begin{document}

\author{ C. Charmousis\thanks{E-mail : christos@celfi.phys.univ-tours.fr}, B. Boisseau\thanks{E-mail : boisseau@celfi.phys.univ-tours.fr},
and B. Linet\thanks{E-mail : linet@celfi.phys.univ-tours.fr} \\
\small Laboratoire de Math\'ematiques et Physique Th\'eorique \\
\small CNRS/UPRES-A 6083, Universit\'e Fran\c{c}ois Rabelais \\
\small Facult\'e des Sciences et Techniques \\
\small Parc de Grandmont 37200 TOURS, France }

\title{\bf Dynamics of a self-gravitating  global string with large Higgs boson mass}

\date{}
\maketitle

\begin{abstract}

We consider a self-gravitating  string generated by a global vortex solution in general relativity. We investigate the Einstein and field equations of a global vortex in the region of its central line and at a distance from the centre of the order of the inverse of its Higgs boson mass. By combining the two we establish  by a limiting process of large Higgs mass the dynamics of a self-gravitating global string. Under our assumptions the presence of gravitation restricts the world sheet of the global string to be totally geodesic.     

{\em PCAS numbers :  11.27.+d, 04.40.-b}
\end{abstract}

\thispagestyle{empty}

\newpage
\renewcommand{\thesection}{\Roman{section}}

\section{Introduction}

 It is generally admitted that a cosmic string obeys the Nambu-Goto dynamics in curved background spacetime. Indeed this has been shown for a gauge string embedded in Minkowski space-time \cite{Nielsen-Olesen, Forster} in the limit where the range of the massive fields tends to zero. In general relativity, studies on infinitely thin straight cosmic strings \cite{Vilenkin} establish their self-gravitating nature : they are  singular lines whose points are conical singularities. In generalising this for curved self-gravitating cosmic strings several  authors stated \cite{vic, fro, unr, cla, isr2} that they generate totally geodesic world sheets which are very particular cases of the history of a Nambu-Goto string.

In a recent work \cite{bcl} this paradoxical property has been raised by considering a geometrical model  of a tube of matter. Under a specific constraint, it is possible for the world sheet of the central line to really obey  the Nambu-Goto dynamics in the limit when its radius goes to zero. The question that naturally arises is  what happens for a field theoretical model of matter. 

The purpose of this work is to examine in the context of general relativity the dynamics of a self-gravitating  string generated by matter,
 described by a complex scalar field $\Phi$ embedded in a "Mexican hat" 
potential whose Lagrangian is given by,
\begin{equation}
\label{1}  
L=-\frac{1}{2}g^{\small{\mu \nu}}\partial_{\small{\mu}}\Phi
\partial_{\small{\nu}}\overline{\Phi}-\frac{\lambda}{8}(|\Phi|^{2}-\eta^{2})^{2},
\end{equation} 
where $\lambda$ and $\eta$ are positive coupling constants \footnote{We have taken signature ($-+++$).}. In the following we take without loss of generality $\eta=1$ and so  $\sqrt{\lambda}$ is the Higgs mass.  We consider global vortex solutions of (\ref{1}), because it is the most simple case of a  vortex solution in field theory, involving only a scalar complex field.

We seek solutions which describe a curved global vortex in the coupled  $\Phi$-field and Einstein equations,
\begin{equation}
\label{2}  
\Box\Phi-\frac{\lambda}{2}\Phi(|\Phi|^2-1)=0,
\end{equation}
\begin{equation}
\label{3}  
R_{\small{\mu\nu}}=8 \pi G(T_{\small{\mu\nu}}-\frac{1}{2}g_{\small{\mu\nu}}T),
\end{equation}
with the energy-momentum tensor $T_{\mu\nu}$ derived from (\ref{1}), 
\begin{equation}
\label{4}  
T_{\alpha\beta}=\frac{1}{2}(\partial_\alpha\overline{\Phi}\partial_\beta\Phi+\partial_\alpha\Phi\partial_\beta\overline{\Phi})-g_{\alpha\beta}\left[\frac{1}{2}\partial_\gamma\overline{\Phi}\partial^\gamma\Phi+\frac{\lambda}{8}(|\Phi|^2-1)^2\right].
\end{equation}
To our knowledge, this problem has never been analytically studied.

Let us define the width parameter $\epsilon$ as being the inverse of the Higgs mass, $\epsilon=1/\sqrt{\lambda}$. To avoid confusion we shall name global string a global vortex with $\epsilon$ going to zero.

In the present work we confine ourselves to a class of metrics in order to solve equations (\ref{2}) and (\ref{3}) in which the global vortex approaches a straight one in a small neighborhood. The central line of the vortex, defined by the null value of the field $\Phi$, has an arbitrary smooth shape with a curvature much larger than the width parameter $\epsilon$.

We first determine a general equation satisfied by the central line of the global vortex, independantly of the value of the width parameter $\epsilon$. This generalises to our class of metrics the equation established by 
 Ben-Ya'acov \cite{ben} in the Minkowski space-time for a general vortex solution.

Secondly we  examine the field equations (\ref{2}) and (\ref{3}) at a distance $l_0$ from the central line  of the order  of the width parameter $\epsilon$. In fact $l=l_0$ is just the boundary of the core of the vortex. The method then consists in an expansion of the field equations (\ref{2}) and (\ref{3}) in powers of $1/\epsilon$ at $l=l_0$. We then  take the mathematical limit of the  width parameter $\epsilon$ going to zero keeping the ratio $l_0/\epsilon$ constant, since $l_0$ and $\epsilon$ are of the same order. Let us point out that this limiting process is different from the one described in the previous paragraph in which the central line is obtained by $l\rightarrow0$ for fixed $\epsilon$.
 
For our class of metrics, the $\Phi$-field equation (\ref{2})  in the limit where $\epsilon\rightarrow0$ combined with the previous general equation gives the Nambu-Goto equation for the central line. However, when we consider the Einstein equations, it is proved that the equation of motion for the central line is very restricted when the Higgs boson mass tends to infinity. The world sheet of the self-gravitating global string is totally geodesic. We find then the surprising result that the presence of gravitation does not permit the curving of the global string. 

The paper is organised as follows. In Sec. II we review some of the well known facts about cylindrically symmetric global vortex solutions in general relativity \cite{har, Cohen}. In Sec. III we describe the gravitational and field theoretical model used to describe these curved vortices in general relativity. In Sec. IV we calculate the general equation of motion for the central line of the vortex and we give the Einstein equations for the central line of a vortex. We calculate geometrical and matter quantities in Sec. V at a distance  $l_0$ of order $\epsilon$ from the central line of the vortex, and by a limiting process we find the equations of motion of a self-gravitating global string in Sec. VI. Finally we summarise our results and  give some concluding remarks in Sec. VII.
\newpage

\section{The straight global vortex}

In the coordinate system
 $(t,z,l,\theta)$ with $l\geq0$
 and $0\leq\theta\leq2\pi$ 
we consider a static metric with cylindrical symmetry with respect
 to the $z$-axis and Lorentz invariance with respect to the $(t,z)$-plane,
\begin{equation}
\label{5}  
ds^{2}=A\left(\frac{l}{\epsilon}\right)(-dt^{2}+dz^{2})+dl^{2}
+l^{2}B\left(\frac{l}{\epsilon}\right)d\theta^{2}.
\end{equation}
We  have rescaled the positive functions $A$ and $B$ by the width parameter
 $\epsilon$. Regularity of metric (\ref{5}) at $l=0$ imposes initial conditions on the functions $A$ and $B$ which have the following form near $l=0$
\begin{equation}
\label{6}  
A\left(\frac{l}{\epsilon}\right)=1+\alpha\frac{l^2}{\epsilon^2}+O\left(\frac{l^4}{\epsilon^4}\right)
,\qquad  B\left(\frac{l}{\epsilon}\right)=1+\beta\frac{l^2}{\epsilon^2}+O\left(\frac{l^4}{\epsilon^4}\right),
\end{equation}
with $\alpha$ and $\beta$ constants.

We consider  the $\Phi$-field and Einstein equations (\ref{2}), (\ref{3}) coupled with metric (\ref{5}).
We seek static, cylindrically symmetric  vortex solutions to this system of differential equations; hence solutions where $\Phi$ is in the form
\begin{equation}
\label{7}  
\Phi=\phi(\frac{l}{\epsilon})e^{i\theta},
\end{equation}
with $\phi$ a real function such that
\begin{equation}
\label{8}  
\phi\left(\frac{l}{\epsilon}\right)=\frac{l}{\epsilon}\left(c+d\frac{l^2}{\epsilon^2}+O\left(\frac{l^4}{\epsilon^4}\right)\right),\quad\mbox{as}\quad l\longrightarrow 0
\end{equation}
\begin{equation}
\label{9}  
\phi(\frac{l}{\epsilon})\sim 1,\quad\mbox{as}\quad l\longrightarrow \infty\nonumber
\end{equation}
with $c$ and $d$ constants. Asymptotic behavior (\ref{8}) is imposed to ensure the regularity of $\phi$ at $l=0$.
For simplicity  we have taken the winding number of the vortex $\Phi$ equal to 1.

From (\ref{2}) and (\ref{3}) the functions $A$, $B$ and $\phi$ are solutions of the system of
 differential equations, 
\begin{equation}
\label{10}  
\phi''+\left(\frac{A'}{A}+\frac{B'}{2B}+\frac{\epsilon}{l}\right)\phi'-
\frac{\epsilon^2}{l^2B}\phi-\frac{\phi}{2}(\phi^2-1)=0,
\end{equation}
\begin{eqnarray}
\label{11}
\left.
\begin{array}{ccc}  
\displaystyle{A''\strut\over A}+\frac{A'B'}{2AB}+\frac{\epsilon A'}{lA}
=-2\pi G(\phi^2-1)^2,\\  
\displaystyle{{2A''}\strut\over A}+\frac{B''}{B}-\frac{{A'}^2}{A^2}-\frac{{B'}^2}{2B^2}+
\frac{2\epsilon B'}{lB}=-8\pi G\left[\frac{1}{4}(\phi^2-1)^2+2{\phi'}^2\right],\\
\displaystyle{B''\strut\over B}-\frac{{B'}^2}{2B^2}+\frac{A'B'}{AB}+\frac{2\epsilon A'}{lA}+\frac{2\epsilon B'}{lB}=-8\pi G\left[\frac{1}{4}(\phi^2-1)^2+2\frac{\phi^2\epsilon^2}{Bl^2}\right].
\end{array}
\right\}
\end{eqnarray} 
Here prime denotes differentiation with respect to the variable
 $l/\epsilon$.
Initial conditions (\ref{6}), (\ref{8}) and (\ref{9}) imposed along with  the system of non-linear differential equations (\ref{10}), (\ref{11}) determine the functions $A$, $B$ and
 $\phi$ for all values of $l$.

 Another important characteristic of the global vortex which will not occupy us here is their singular nature which generates singular global string spacetimes \cite{ruth, ruiz}. A recent discussion on the singularity of global strings can be found in \cite{greg}.  In what follows we shall always stay in proximity of the core of the vortex where the solution is well defined.

\section{The curved vortex model}

We generalise the metric given in (\ref{5}) and the 
form of the vortex solution (\ref{7}) in order to permit the curving of
 the vortex.  
Let us consider the following class of metrics which we suppose valid in the core neighborhood,
\begin{equation}
\label{12}  
ds^{2}=g_{\small{AB}}(\tau^{\small{A}},l,\theta)
d\tau^{\small{A}}d\tau
^{\small{B}}
+dl^{2}+l^{2}
B\left(\frac{l}{\epsilon}\right)d\theta^{2}
\end{equation}
where 
\[
g_{\small{AB}}(\tau^{\small{A}},l,\theta)=A(\frac{l}{\epsilon})
g^\dagger_{\small{AB}}(\tau^{\small{A}},l,\theta)\] 
with indices  $A,B=0,3$. The functions $A$ and $B$
denote again the straight vortex solutions for metric (\ref{5}). It is also important to note, that $g^\dagger_{\small{AB}}$ varies on a scale larger than the string width parameter $\epsilon$. This permits us to suppose that the smooth components $g^\dagger_{\small{AB}}(\tau^{\small{A}},l,\theta)$ do not depend explicetely on the string width parameter $\epsilon$. This metric is reminiscent to the {\it smooth cone metric} which was introduced in \cite{bcl}, the difference being that here the functions $A$ and $B$ are determined by the matter via equations (\ref{10}), (\ref{11}) and that the parameter $\epsilon$ is a physical parameter. 

We introduce coordinates
 $\rho^{\small{a}}$, ($a=1,2$) which are well defined at $l=0$,
\[  
\rho^{1}=l\cos\theta\qquad \rho^2=l\sin\theta.\nonumber
\]
Then metric (\ref{12}) becomes,
\begin{equation}
\label{14}  
ds^{2}=A(\frac{l}{\epsilon})g_{\small{AB}}^\dagger(\tau^{\small{A}},\rho^a)
d\tau^{\small{A}}d\tau
^{\small{B}}
+g_{ab}(\rho^{a})d\rho^a d\rho^b
\end{equation}
where the components of $g_{ab}$ and their determinant $\stackrel{\wedge}{g}$ are given by the following expression,
\begin{equation}
\label{15}
g_{ab}=\delta_{ab}+q\left(\frac{l}{\epsilon}\right)\varepsilon_{ac}
\varepsilon_{bd}\frac{\rho^{c}\rho^{d}}{\epsilon^{2}},\qquad
\stackrel{\wedge}{g}=\det(g_{ab})=1+\frac{l^{2}}{\epsilon^{2}}
q\left(\frac{l}{\epsilon}\right)
\end{equation}        
where $\varepsilon_{ac}$ is the totally antisymmetric Levi-Civita 
symbol, $\stackrel{\wedge}{(\mbox{ } )}$ stands for the induced 
geometrical quantities of the 2-dimensional surface $\tau^A=$const. and $q$ is a function defined in terms of 
the straight vortex solution  $B$,
\begin{equation}
\label{16}
q\left(\frac{l}{\epsilon}\right)=\frac{\epsilon^2}{l^2}\left(B\left(\frac{l}{\epsilon}\right)-1\right). 
\end{equation}
Coordinates $\rho^a$ turn out to be geodesic coordinates of the spacetime. For convenience we note $\rho_a=g_{ab}\rho^b$ which from the form (\ref{15}) of the $g_{ab}$ gives $\rho_a=\rho^a$.

The metric components $g_{AB}$ can be expanded in powers of $\rho^a$ in the $l=0$ vicinity of the vortex,
\begin{equation}
\label{18}
g_{AB}(\tau^A,\rho^a)=\gamma_{AB}+2K_{aAB}\rho^a+[K_{bA}^DK_{aBD}-(R_{AbBa})_0]\rho^a\rho^b+O(l^3), 
\end{equation}
where $(R_{AbBa})_0=R_{AbBa}(\tau^A,0)$. We note by $\gamma_{AB}$ and $K_{aAB}$ the induced metric and  the extrinsic curvature of the world sheet $l=0$ respectively. In fact bearing in mind, the asymptotic form (\ref{6}) of  function $A$ we get
\begin{equation}
\label{17}
\gamma_{AB}(\tau^A)=g_{AB}^\dagger(\tau^A, \rho^a)_{|l=0}, 
\end{equation}
\begin{equation}
\label{19}
K_{aAB}(\tau^A)=\frac{1}{2}\partial_ag_{AB}^\dagger(\tau^A, \rho^a)_{|l=0}. 
\end{equation}

On the other hand we now consider that  the scalar complex field  $\Phi$ describes a curved vortex solution in the form : 
\begin{equation}
\label{20}
\Phi(\tau^A,\rho^a)=\phi(\frac{l}{\epsilon})e^{i\theta}\psi(\tau^A,\rho^a).
\end{equation}
In the above expression $\phi$ is the straight vortex solution described in the previous 
section and $\psi$ is an unknown function, which describes the curvature corrections imposed by the curving of the global vortex. 

The closer we are to the central line $l=0$ the more the vortex $\Phi$ resembles the straight vortex $\phi e^{i\theta}$. It is plausible then to suppose that the correction function $\psi$ is analytic in the vortex interior $(l\leq \epsilon)$ and non null on the central line $l=0$. In fact since $\phi(0)=0$ we can take for instance $\psi(\tau^A,0)=1$.

\section{Equation of the central line of the vortex}

By using  definition (\ref{20}) the equation of the central line of a general vortex of arbitrary shape has been found in the Minkowski spacetime \cite{ben}.  We shall follow this method to determine the equation of the central line of a vortex of arbitrary shape embedded in a space-time described by a class of metrics (\ref{14}).
In order to do this we shall calculate the $l\rightarrow0$ limit of the $\Phi$-field equation (\ref{2}) leaving the width parameter $\epsilon$ fixed.

Taking into account (\ref{8}) our vortex solution  becomes,
\begin{equation}
\label{21}
\Phi(\tau^A,\rho^a)=\frac{l}{\epsilon}\left(c+d\frac{l^2}{\epsilon^2}+O(\frac{l^4}{\epsilon^4})\right)e^{i\theta}\psi(\tau^A,\rho^a)\quad\mbox{as}\quad l\rightarrow0.
\end{equation}
On the other hand the d'Alembertian of $\Phi$ for metric (\ref{14}) can be expanded in the following way :
\begin{eqnarray}
\label{22}
\Box\Phi&=&\frac{1}{\sqrt{\stackrel{\wedge}{g}}}\partial_a(g^{ab}\sqrt{\stackrel{\wedge}{g}})\partial_b\Phi+g^{ab}\partial_a(\ln\sqrt{-g^*})\partial_b\Phi\nonumber\\
&+&g^{ab}\partial_a\partial_b\Phi+\frac{1}{\sqrt{-g^*}}\partial_A(\sqrt{-g^*}g^{AB}\partial_B\Phi),
\end{eqnarray}
where $g^*$ is the determinant of the $g_{AB}$ components.
 We insert  solution (\ref{21}) in expression (\ref{22}). Taking the $\Phi$-field equation (\ref{2}) at the  $l\rightarrow0$ limit the potential term  and  
the last term in (\ref{22}) drop out and  using (\ref{6}) we obtain,
\begin{equation}
\label{24}
(\frac{\partial}{\partial\rho^1}+i\frac{\partial}{\partial\rho^2})\left(2\ln\psi+\ln\sqrt{-g^\dagger}\right)_{|l=0}=0.
\end{equation}
 By (\ref{19}) we have that 
\begin{equation}
\label{25}
\left(\frac{\partial}{\partial{\rho^a}}\ln \sqrt{-g^\dagger}\right)_{|l=0}=K_a,
\end{equation}
where $K_a=K_{aAB}\gamma^{AB}$ is the mean curvature of the world sheet $l=0$ and by setting, 
\begin{equation}
\label{26}
\ln\psi(\tau^A,\rho^a)=\mu(\tau^A,\rho^a)+i\nu(\tau^A,\rho^a)
\end{equation}
with $\mu$ and $\nu$ real functions we obtain from (\ref{24}) the final form of the equation for the centre of the vortex,
\begin{equation}
\label{27}
K_a(\tau^A)+2\left[\partial_a\mu(\tau^A,0)-\varepsilon_{ab}\partial_b\nu(\tau^A,0)\right]=0.
\end{equation}
 We note that the above equation, valid for arbitrary $\epsilon$ is an exact equation for the central line of a vortex and is identical with that found in \cite{ben} for flat spacetime. It gives the relation between the mean curvature $K_a$ and the field of the global vortex.

 For completeness since we are in the presence of gravity we must also consider the Einstein equations for $l\rightarrow0$.
Indeed on using (\ref{17}), (\ref{19}) and  on considering that the functions $A$ and $B$ in (\ref{6}) are solutions of the straight vortex Einstein equations (\ref{11}) we obtain for $l=0$ : 
\begin{equation}
\label{28}
(R^*_{AB})_{|l=0}-K_aK_{AB}^{a}+2K_{aA}^{D}K^{a}_{BD}
-\frac{1}{2}(\partial^a\partial_ag_{AB}^\dagger)_{|l=0}=0
\end{equation}
\begin{equation}
\label{29}
\partial_BK_a=\bigtriangledown^*_A K_{aB}^{A},
\end{equation}
\begin{equation}
\label{30}
\gamma^{AB}(\partial_a\partial_b g_{AB}^{\dagger})_{|l=0}=2K_{aAB}K_b^{AB}
\end{equation}
where $(\mbox{ })^*$ stands for geometrical quantities issued from the metric $g_{AB}$ which in this case are just the induced quantities on the world sheet $l=0$.
 Let us note that equation (\ref{27}) gives a relation between the geometry of the world sheet and the correction function of the field whereas the Einstein equations (\ref{28})-(\ref{30}) can be interpreted as constraint equations on the extrinsic curvature of the world sheet.

Indeed constraint equations are also obtained  for a curved global vortex embedded in the Minkowski spacetime. We remark that the vortex we have considered has no twist since we have omitted cross terms in metric (\ref{14}). 
In this simple case equations (\ref{28})-(\ref{30}) are just the Gauss-Codazzi integrability conditions for flat spacetime (see for example \cite{mac, cap}) :
$$
R^*_{AB}-K_aK_{AB}^{a}+K_{aA}^{D}K^{a}_{BD}=0,\qquad
\partial_BK_d-\bigtriangledown^*_AK_{aB}^{A}=0
$$
where equation (\ref{30}) is identically verified.

\section{Expansion of geometrical and matter \\ quantities}

 In the sections that follow we shall establish an equation of motion for a curved global string, that is for a vortex with width parameter $\epsilon$ going to $0$. It is evident that the width parameter $\epsilon$ must play a leading role in this process. We shall now place ourselves in the vicinity of the boundary of the vortex core  considering that the distance  $l=l_0$ from the central line of the vortex is of the same order as $\epsilon$. In consequence when we shall vary $\epsilon$, the distance $l_0$ shall vary in the same way so that : 
\[\frac{l_{\small{0}}}{\epsilon}=\mbox{const.}
\]
In this section we shall proceed in the manner of \cite{bcl} in classifying in powers of $\epsilon$ geometrical and matter quantities. It is very important to note that $l_{\small{0}}$, $\epsilon$ or coordinates $\rho^a_{\small{0}}$ on the core boundary are now of the same order, whereas the polar angle $\theta$ remains arbitrary.

Let us first of all remark that the straight vortex solutions are of $0$-order in $\epsilon$,
\[
A\left(\frac{l_{\small{0}}}{\epsilon}\right)=O(\frac{1}{\epsilon^0}),\qquad B\left(\frac{l_{\small{0}}}{\epsilon}\right)=O(\frac{1}{\epsilon^0}),\qquad\phi\left(\frac{l_{\small{0}}}{\epsilon}\right)=O(\frac{1}{\epsilon^0}).
\]
since they only depend on the quotient $l_{\small{0}}/\epsilon$.
Let us also point out  that the straight vortex solutions $A(l_0/\epsilon)$, $B(l_0/\epsilon)$, $\phi(l_0/\epsilon)$ have no longer the $l\rightarrow0$ asymptotic behavior (\ref{6}) and (\ref{8}) because although $l$ varies very little between $0$ and $l_0$, the functions $A$, $B$ and  $\phi$ can vary essentially in this interval.

The metric components given by spacetime (\ref{14}) are of $0$-order,
$$
(g_{ab})_{\small{l=l_{\small{0}}}}=O(\frac{1}{\epsilon^0}),\qquad (g_{AB})
_{l=l_0}=O(\frac{1}{\epsilon^0});
$$
the same holding for their inverse and the determinant, $g=\det g_{\alpha\beta}$.
We can calculate the non-null Christoffel coefficients for $l=l_{\small{0}}$,
$$
\Gamma_{BC}^{A}=\Gamma_{BC}^{A*}=O(\frac{1}{\epsilon^0}),
$$
$$
\Gamma_{aB}^{D}=\frac{1}{2}g^{AD}g_{AB,a}=\frac{1}{2\epsilon}\frac{A'}{A}\delta_B^D\frac{\rho_{a0}}{l_0}+\frac{1}{2}g^{\dagger AD}g_{AB,a}^\dagger=O(\frac{1}{\epsilon}),
$$
$$
\Gamma_{AB}^{a}=-\frac{1}{2}g^{ab}g_{AB,b}=-\frac{1}{2\epsilon}A'g_{AB}^\dagger \frac{\rho^a_0}{l_0}-\frac{1}{2}Ag^{ab}g_{AB,b}^\dagger=O(\frac{1}{\epsilon}),
$$
$$
\Gamma_{ab}^{c}=
\stackrel{\wedge}{\Gamma}_{ab}^{c}=O(\frac{1}{\epsilon}).
$$
We recall that $(\mbox{ })^*$ and $\stackrel{\wedge}{(\mbox{ } )}$ stand for quantities derived from $g_{AB}$ and $g_{ab}$ respectively. Notice also, that $\dagger$-terms are always of $0$-order in $1/\epsilon$. 

We now proceed in calculating the Ricci tensor for $l=l_{\small{0}}$. Since this involves the second derivatives of the metric we shall obtain terms up to the second order in $1/\epsilon$,
\begin{equation}
\label{49}
R_{\alpha\beta}\equiv R_{\alpha\beta}(\frac{1}{\epsilon^2})+R_{\alpha\beta}(\frac{1}{\epsilon})+R_{\alpha\beta}(\frac{1}{\epsilon^0}).
\end{equation} 
Therefore for the capital indices we have, 
\begin{equation}
\label{50}
R_{AB}(\frac{1}{\epsilon^2})=-\frac{g_{AB}^\dagger}{2\epsilon^2}
\left[A''+A'\frac{l_{\small{0}}}{\epsilon}
\frac{q+q'\frac{l_{\small{0}}}{2\epsilon}}{B}+A'\frac{\epsilon}{l_{\small{0}}}\right],
\end{equation}
and
\begin{equation}
\label{51}
R_{AB}(\frac{1}{\epsilon})=-\frac{1}{2\epsilon}\frac{\rho^a_{\small{0}}}{l_{\small{0}}}\left[X g_{AB,a}^\dagger +\frac{1}{2}A'g_{AB}^\dagger g^{\dagger CD}g_{CD,a}^\dagger  \right],
\end{equation}
where we have put,
\begin{equation}
\label{52}
X=A'+\frac{l_0}{\epsilon}A\frac{2q+q'\frac{l_{\small{0}}}{2\epsilon}}{B}.
\end{equation}
We do not give the explicit form  of $0$-order terms in $1/\epsilon$ since they  will not be needed in the following.
The cross terms are of zero order in $1/\epsilon$ 
\begin{equation}
\label{53}
R_{aB}=R_{aB}(\frac{1}{\epsilon^0}).
\end{equation}
Finally for the small case indices we have for $l=l_0$ :
\begin{equation}
\label{55}
R_{ab}(\frac{1}{\epsilon^{2}})=\stackrel{\wedge}{R}_{ab}
+\frac{\partial_dA}{A}
\stackrel{\wedge}{\Gamma}_{ab}^{d}
-\frac{\partial_a\partial_bA}{A}
+\frac{1}{2}\frac{\partial_aA\partial_bA}{A^2}
\end{equation}
and
\begin{equation}
\label{56}
R_{ab}(\frac{1}{\epsilon})=\left(\frac{1}{2}\stackrel{\wedge}{\Gamma}_{ab}^{d}-\frac{1}{4}\frac{\partial_bA}{A}\delta_a^d-\frac{1}{4}\frac{\partial_aA}{A}\delta_b^d\right)g^{\dagger CD}g^\dagger_{CD,d}.
\end{equation}

We note that  $R_{AB}(1/\epsilon^2)$ and $R_{ab}(1/\epsilon^2)$ are just the non-zero components of the Ricci tensor for the straight vortex (cf. Sec. II), multiplied by $g_{AB}^\dagger$ and $\delta_{ab}$  respectively. This fact will be important for the next section. 

Let us now proceed in the same way in order to classify matter.
Our vortex solution (\ref{20}) is now written for $l=l_0$:
$$
\Phi(\tau^A,\rho^a)_{|l=l_{\small{0}}}=\phi(\frac{l_{\small{0}}}{\epsilon})e^{i\theta}\psi(\tau^A,\rho^a_0)
$$
and so is of $0$-order in $1/\epsilon$. The first and second derivatives of the straight vortex are respectively  of first and second order in $1/\epsilon$,
$$
(\partial_a\phi e^{i\theta})_{|{l=l_{\small{0}}}}=O(\frac{1}{\epsilon})\qquad
(\partial_a\partial_b\phi e^{i\theta})_{|l=l_{\small{0}}}=O(\frac{1}{\epsilon^2})
$$
whereas the derivatives of $\psi$ are of 0-order in $1/\epsilon$.

We can classify the energy-momentum tensor $T_{\alpha\beta}$ given in (\ref{4}) in powers of  $1/\epsilon$ for $l=l_{\small{0}}$. We note that the potential is of order $1/\epsilon^2$ since $\lambda=1/\epsilon^2$. We have :
\begin{equation}
\label{59}
T_{\alpha\beta}\equiv T_{\alpha\beta}(\frac{1}{\epsilon^2})+T_{\alpha\beta}(\frac{1}{\epsilon})+T_{\alpha\beta}(\frac{1}{\epsilon^0}).
\end{equation}
So we get
\begin{equation}
\label{60}
T_{AB}(\frac{1}{\epsilon^2})=
-\frac{1}{2\epsilon^2}
Ag_{AB}^\dagger
\left[|\psi|^2\left(
\phi'^2
+\frac{\epsilon^2}{l_{\small{0}}^2}
\frac{\phi^2}{B}\right)+\frac{1}{4}(\phi^2|\psi|^2-1)^2\right]
\end{equation}
and
\begin{equation}
\label{61}
T_{AB}(\frac{1}{\epsilon})=-\frac{1}{\epsilon}A g_{AB}^\dagger
\phi(\frac{l_{\small{0}}}{\epsilon})
|\psi|^2
\Re e[\Pi^a\partial_a\ln\psi]_{|l=l_{\small{0}}}.
\end{equation}
where we have set,
$$
\Pi^a=\phi'\frac{\rho^a}{l}+i\phi\frac{\epsilon \varepsilon^{ac}\rho_c}
{Bl^2}.
$$
In the same way we have :
 \begin{eqnarray}
\label{64}
T_{ab}(\frac{1}{\epsilon^2})=
&+&\frac{1}{\epsilon^2} \left\{ |\psi|^2
\left( \frac{\rho_{a0}\rho_{b0}}{{l_0}^2}
\phi'^2
+\frac{\epsilon^2}{l_{\small{0}}^2}
\frac{\varepsilon_{ac}\varepsilon_{bd}\rho^{c}_0\rho^{d}_0}
{l_{\small{0}}^2}\phi^2 \right) \right.\nonumber\\&-&
\frac{1}{2}g_{ab}\left[|\psi|^2\left(\phi'^2
+\frac{\epsilon^2}{l_0^2}
\frac{\phi^2}{B} \right)+\left.\frac{1}{4}(\phi^2|\psi|^2-1)^2\right]\right\} .
\end{eqnarray}
and 
\begin{eqnarray}
\label{65}
T_{ab}(\frac{1}{\epsilon})&=&\frac{1}{2\epsilon}
\phi|\psi|^2 \left[\Re e(\Pi_a\partial_b\ln\psi)_{|l=l_0}+\Re e(\Pi_b\partial_a\ln\psi)_{|l=l_0} \right.\nonumber\\&-&\left.2g_{ab}\Re e(\Pi^d\partial_d\ln\psi)_{|l=l_0} \right]
\end{eqnarray}
where now,  
\begin{eqnarray*}
\Pi_a=\phi'\frac{\rho_a}{l}+i\phi\frac{\epsilon \varepsilon_{ac}\rho^c}
{l^2}.
\end{eqnarray*}
For the mixed indices we have that the energy-momentum tensor is of order $1/\epsilon$ 
\begin{equation}
\label{67}
T_{aA}(\frac{1}{\epsilon})= \frac{1}{\epsilon}\phi|\psi|^2\Re e(\Pi_a\partial_A\ln\Psi)_{|l=l_0}.
\end{equation}
The trace of the energy-momentum tensor induced from the above is of the form,
\begin{equation}
\label{68}
T\equiv T(\frac{1}{\epsilon^2})+T(\frac{1}{\epsilon})+T(\frac{1}{\epsilon^0})
\end{equation}
with
\begin{equation}
\label{69}
T(\frac{1}{\epsilon^2})=-\frac{1}{\epsilon^2}\left[|\psi|^2\left(
\phi'^2+\frac{\epsilon^2}{l_0^2}
\frac{\phi^2}{B}\right)+\frac{1}{2}(\phi^2|\psi|^2-1)^2\right]
\end{equation}
and
\begin{equation}
\label{70}
T(\frac{1}{\epsilon})=-\frac{2}{\epsilon}\phi|\psi|^2\Re e(\Pi^a\partial_a\ln\Psi)_{|l=l_0}.
\end{equation}

As for the $\Phi$-field equation we need to calculate the d'Alembertian of $\Phi$ for $l=l_0$,
\begin{equation}
\label{71}
\Box\Phi\equiv\Box\Phi(\frac{1}{\epsilon^2})+\Box\Phi(\frac{1}{\epsilon})+\Box\Phi(\frac{1}{\epsilon^0}).
\end{equation}
We shall again be interested in the first two terms of the expansion,
\begin{equation}
\label{72}
\Box\Phi(\frac{1}{\epsilon^2})=\psi\Box_{str.}(\phi e^{i\theta})
u\end{equation}
where $\Box_{str.}$ is the d'Alembertian operator of metric (\ref{5}) and
\begin{equation}
\label{73}
\Box\Phi(\frac{1}{\epsilon})=\frac{1}{\epsilon}
\left[\frac{X}{A}\phi
\frac{\rho_{0}^a}{l_0}\partial_a(\ln\psi)_{l=l_0}+
\overline{\Pi}^{a}\partial_a(\ln \sqrt{-g^\dagger}+2\ln\psi)_{l=l_0}\right]e^{i\theta}\psi.
\end{equation}
Note that the function $X$ given in (\ref{52}) reappears in the $\Phi$-field equation.

\section{Equations of motion for the global string}

We consider the coupled system of the $\Phi$-field and Einstein equations  (\ref{2}) and (\ref{3}) for $l=l_0$.
Let us first examine the $\Phi$-field equation :
\begin{equation}
\label{74}
\Box\Phi(\frac{1}{\epsilon^2})+\Box\Phi(\frac{1}{\epsilon})+\Box\Phi(\frac{1}{\epsilon^0})-\frac{\Phi}{2\epsilon^2}(|\Phi|^2-1)=0.
\end{equation}
 We focus our attention to the correction function $\psi$. As we mentioned in Sec. III the correction function is supposed analytic for $0\leq l\leq l_0$ and so can be expanded around $l=0$. This reflects the fact that unlike the straight vortex solution $\phi$, the correction $\psi$ varies on a larger scale so can be evaluated on $l=l_0$ by its values on $l=0$ just like the metric components $g_{AB}^\dagger$ in (\ref{18}). We have,
\begin{equation}
\label{76}
\psi_{|l=l_0}=1+(\partial_z\psi)_{|l=0}z_0+O(l^2),
\end{equation}
where $z=\rho_1+i\rho_2$.
So the potential term is written,
\begin{equation}
\label{75}
-\frac{\Phi}{2\epsilon^2}(|\Phi|^2-1)=-\frac{1}{2\epsilon^2}(\phi^2-1)\phi e^{i\theta}\psi-\frac{\phi^3e^{i\theta}\psi}{\epsilon}\Re e\left[(\partial_z\psi)_{|l=0}\frac{z_0}{\epsilon}\right].
\end{equation}
Introducing (\ref{75}) in (\ref{74}) and using (\ref{72}), the $1/\epsilon^2$ terms cancel out since they are issued from the straight vortex equations (\ref{10}).
This leaves the $1/\epsilon$ as leading terms in the $\Phi$-field equation (\ref{74}) which can now be written for $l=l_0$,
\begin{eqnarray}
\label{78}
&1/\epsilon&\left[\frac{X}{A}\phi\frac{\rho_{0}^a}{l_0}(\partial_a\ln\psi)_{|l=l_0}+
\overline{\Pi}^{a}[\partial_a(\ln \sqrt{-g^\dagger}+2\ln\psi)]_{|l=l_0}\right]e^{i\theta}\psi\nonumber\\
-&1/\epsilon&\left\{\phi^3\Re e\left[(\partial_z\psi)_{|l=0}\frac{z_0}{\epsilon}\right]\right\}\psi e^{i\theta}+0(\frac{1}{\epsilon^0})=0.
\end{eqnarray}
On writing the correction function $\psi$ in form (\ref{26}) we split up the above equation in its real and imaginary part. We then let $\epsilon\rightarrow0$ and since
$$
\partial_a(\ln\sqrt{-g^\dagger})_{|l=0}=\lim_{\epsilon\rightarrow0}(\partial_a\ln\sqrt{-g^\dagger})_{|l=l_0}
$$
we get inputing the mean curvature by formula (\ref{25}), 
\begin{eqnarray}
\label{79}
&&\frac{X(\frac{l_0}{\epsilon})}{A(\frac{l_0}{\epsilon})}\phi(\frac{l_0}{\epsilon})\frac{\rho_{0}^a}{l_0}(\partial_a\mu)_{|l=0}
-\phi^3(\frac{l_0}{\epsilon})\Re e\left[(\partial_z\psi)_{|l=0}\frac{z_0}{\epsilon}\right]\\
&&+\phi'(\frac{l_0}{\epsilon})\frac{\rho^a_0}{l_0}(K_a+2\partial_a\mu)_{|l=0}
+\frac{\epsilon\phi(\frac{l_0}{\epsilon})}{l_0B}\left[\frac{\rho_0^2}{l_0}(\partial_1\nu)_{|l=0}-\frac{\rho_0^1}{l_0}(\partial_2\nu)_{|l=0}\right]=0\nonumber
\end{eqnarray}
and
\begin{eqnarray}
\label{80}
&&+\left[\frac{X(\frac{l_0}{\epsilon})}{A(\frac{l_0}{\epsilon})}\phi(\frac{l_{\small{0}}}{\epsilon})+2\phi'(\frac{l_0}{\epsilon})\right]
\frac{\rho_{0}^a}{l_0}(\partial_a\nu)_{|l=0}\nonumber\\&&-
\frac{\epsilon\phi(\frac{l_0}{\epsilon})}{l_0B}
\left[\frac{\rho_0^2}{l_0}(K_1+2\partial_1\mu)_{|l=0}-\frac{\rho_0^1}{l_0}(K_2+2\partial_2\mu)_{|l=0}\right]=0.
\end{eqnarray}
We now make use of the equation (\ref{27}) for the central line of the vortex , and so equation (\ref{80}) gives :
\begin{equation}
\label{81}
\left[\frac{X(\frac{l_{\small{0}}}{\epsilon})}{A(\frac{l_{\small{0}}}{\epsilon})}\phi(\frac{l_{\small{0}}}{\epsilon})+2\phi'(\frac{l_{\small{0}}}{\epsilon})-\frac{\epsilon\phi(\frac{l_{\small{0}}}{\epsilon})}{l_0 B}\right]\frac{\rho_{0}^a}{l_0}(\partial_a\nu)_{|l=0}=0
\end{equation}
which is valid for all $\rho_0^a/l_0$ depending on an arbitrary angle. Hence 
\begin{equation}
\label{82}
(\partial_a\nu)_{|l=0}=0.
\end{equation}
Inserting the above in (\ref{79}) we find in the same way that 
\begin{equation}
\label{83}
(\partial_a\mu)_{|l=0}=0.
\end{equation}
So in the limiting case of a string where $\epsilon\rightarrow0$ the general equations (\ref{27}) reduce to the Nambu-Goto equations of motion,
\begin{equation}
\label{84}
K_a=0.
\end{equation}

Since we are in the presence of gravitation we must also examine the Einstein equations for $l=l_0$. First of all the cross-term Einstein equations give us :
\begin{equation}
\label{85}
R_{aA}(\frac{1}{\epsilon^0})=8\pi G\left(T_{aA}(\frac{1}{\epsilon})+T_{aA}(\frac{1}{\epsilon^0})\right).
\end{equation}
Hence from (\ref{67}) when $\epsilon\rightarrow0$ we get 
\begin{equation}
\label{86}
\Re e(\Pi_a\partial_A\ln\psi)_{|l=l_0}=0,
\end{equation}
and on replacing $\psi$ from (\ref{26}) it is easy to see that 
\begin{equation}
\label{87}
(\partial_A\nu)_{|l=0}=(\partial_A\mu)_{|l=0}=0.
\end{equation}
Hence the correction functions  $\mu$ and $\nu$ are  constant when the width parameter $\epsilon$ goes to zero.

For the $(A,B)$ and $(a,b)$ Einstein equations the $1/\epsilon^2$ terms cancel out from the straight string Einstein equations (\ref{11}). This leaves us just as for the $\Phi$-field equation with the $1/\epsilon$ as leading terms in the expansion of the Einstein equations. Furthermore in the limit when $\epsilon\rightarrow0$ we have found in the previous paragraph that the correction function  $\psi$ is constant and thus the energy-momentum components of order $1/\epsilon$ (\ref{61}) and (\ref{65}) drop out. Hence in the limiting case, the  Einstein equations multiplied by  $\epsilon$ simply  give 
\begin{equation}
\label{88}
\lim_{\epsilon\rightarrow0}\epsilon R_{AB}(\frac{1}{\epsilon})=0,\qquad\lim_{\epsilon\rightarrow0}\epsilon R_{AB}(\frac{1}{\epsilon})=0.
\end{equation}
We have that (\ref{17}) and (\ref{19}) can be also written
\[\gamma_{AB}=\lim_{\epsilon \rightarrow 0}g_{AB}^\dagger(\tau^A,\rho_0^a),\qquad K_{aAB}=\frac{1}{2}\lim_{\epsilon \rightarrow 0}\partial_ag_{AB}^\dagger(\tau^A,\rho_0^a).
\]
 Therefore on inserting the above in  (\ref{51}) and (\ref{56}) the Einstein equations give,
\begin{equation}
\label{89}
\frac{\rho^a_{\small{0}}}{l_{\small{0}}}\left[X\left(\frac{l_{\small{0}}}{\epsilon}\right) K_{aAB}+\frac{1}{2}A'\left(\frac{l_{\small{0}}}{\epsilon}\right)\gamma_{AB}K_a\right]=0    
\end{equation}
\begin{equation}
\label{90}
F_{ab}^{d}\left(\frac{\rho_0^a}{\epsilon}\right)K_d=0,
\end{equation}
where 
\[
F_{ab}^{d}=\lim_{\epsilon \rightarrow 0}\epsilon\left(\stackrel{\wedge}{\Gamma}_{ab}^{d}-\frac{1}{2}\frac{\partial_bA}{A}\delta_a^d-\frac{1}{2}\frac{\partial_aA}{A}\delta_b^d\right)
\]
is a coefficient depending uniquely on the polar angle.
Now the Nambu-Goto equation of motion (\ref{84}), $K_a=0$, annihilates equation (\ref{90}) but not (\ref{89}). In fact in order for (\ref{89}) to be verified either the global string must be totally geodesic $K_{aAB}=0$, or the coefficient of the extrinsic curvature, the function $X$ given by (\ref{52}), must be null at $l=l_0$.

The function $X$ is given in terms of the straight vortex solution so it is totally determined. By means of a graphic numerical solution we can search the zeros, if any, of this function.  In the following figure obtained by numerically solving the straight vortex equations (\ref{10}) and (\ref{11}) we see that the function $X$ is strictly negative except for the flat case which corresponds to the origin ($A=1$, $B=1$). So in the presence of gravitation when the width parameter $\epsilon\rightarrow0$ the global string is totally geodesic, that is $K_{aAB}=0$.

\section{Conclusion}

With a reasonably general form of the metric (\ref{14}) describing a self-gravitating global vortex having an arbitrary shape, we have obtained results concerning the equations of motion of the central line of the global string.

In the first part of this work, we study a global vortex of width parameter $\epsilon=1/\sqrt{\lambda}$. We find a  general equation (\ref{27}) between the mean curvature $K_a$ and the field of the global vortex for the world sheet of the central line, $l=0$. Constraint equations (\ref{28}-\ref{30}) for the extrinsic curvature of the world sheet are issued from the Einstein equations.

In the second part, we consider the $\Phi$-field and Einstein equations at a distance from the centre of the vortex of order $\epsilon$. We make the assumption that the Higgs boson mass is large, i.e. $\epsilon\rightarrow0$. From the $\Phi$-field equations, we deduce (\ref{79}) and (\ref{80}). Moreover by using (\ref{27}) we obtain the Nambu-Goto equation of motion $K_a=0$ for the global string. However in the presence of gravitation the Einstein equations give (\ref{89}) and (\ref{90}) which restrain the global string to be totally geodesic, i.e. $K_{aAB}=0$. So the presence of gravitation does not permit the curving of the global string.

In \cite{bcl} using as a string model a thin tube of undetermined matter we have found a condition for which the self-gravitating string has the Nambu-Goto dynamics. However this condition corresponding here to $X=0$ cannot be fulfilled in the presence of gravitation for  the case of a global string.

It has been shown, \cite{vil, she, ben3, bat} in Minkowsky space-time that the equation of motion of a global string is the Nambu-Goto equation corrected by a term resulting from the radiative backreaction. For example in the review \cite{rep} the authors write in the conformal gauge the following equation of motion (cf. eq. (75)) :
\begin{equation}
\label{*}
(\ddot {x}^\mu-x''^\mu)=\frac{f^\mu_{rad}}{\mu(\Delta)}
\end{equation}
with
$$
\mu(\Delta)=\mu_0+2\pi \ln(\sqrt{\lambda}\Delta),
$$
where $\Delta$ is a fixed renormalisation scale and $\mu_0$ is the energy integral within the core of the global vortex. When the Higgs mass is large we see that the radiative part becomes negligeable. Equation (\ref{*}) becomes the Nambu-Goto equation of motion, $K_a=0$ which is in accord with our result in the absence of gravitation.

\section*{Acknowlegments}
The authors would like to thank U. Ben-Ya'acov for useful discussions in vortex dynamics during his visit in Tours. We would also like to thank B. Carter for bringing to our attention difficulties concerning global strings. Finally we thank S. Neukirch for helping out with the numerical analysis.

\newpage

\begin{figure}[htbp]
\epsfxsize=11cm
$$
\epsfbox{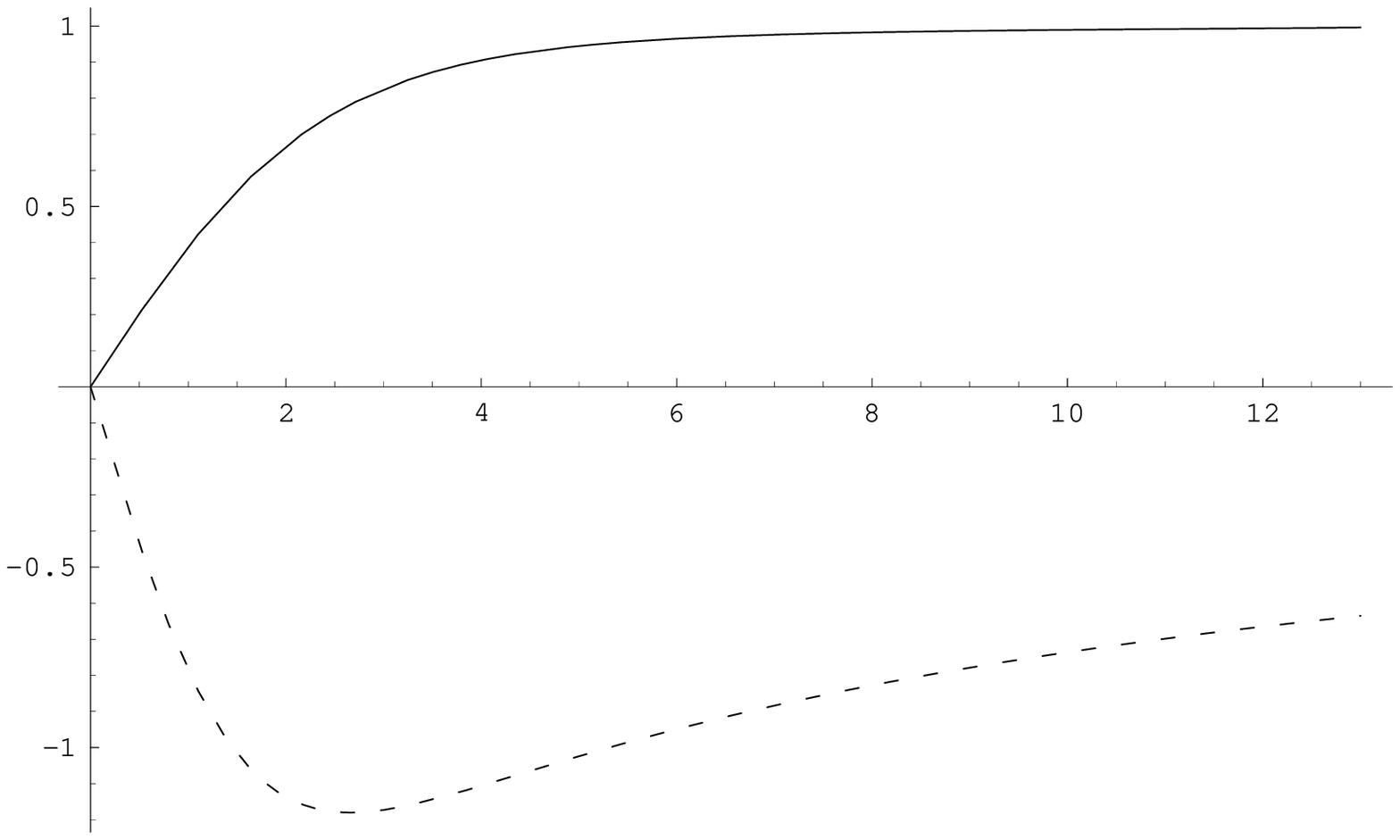}
$$
\caption{The straight global vortex solution $\phi(l/\epsilon)$ (plain line) and function $X(l/\epsilon)$ (dashed line) traced on the same scale in general relativity.}
\end{figure}

\end{document}